\documentclass[%
reprint,
superscriptaddress,
amsmath,amssymb,
prstper,
longbibliography
]{revtex4-1}

\usepackage{color}
\usepackage{graphicx}
\usepackage{dcolumn}
\usepackage{bm}

\begin{document}
\title{Full-wave analysis of reverse saturable absorption in time-domain}
\author{Shaimaa I. Azzam}
\email{sazzam@purdue.edu}
\affiliation{School of Electrical and Computer Engineering and Birck Nanotechnology Center, Purdue University, West Lafayette, IN 47907, USA}
\author{Alexander V. Kildishev}
\email{kildishev@purdue.edu}
\affiliation{School of Electrical and Computer Engineering and Birck Nanotechnology Center, Purdue University, West Lafayette, IN 47907, USA}

\begin{abstract}
An advanced full-wave time-domain numerical model for reverse saturable absorption (RSA) is presented and verified against established methods. Rate equations, describing atomic relaxations and excitation dynamics, are coupled to Maxwell equations by using a Lorentzian oscillator. These oscillator models the kinetics-dependent light-matter interaction in the form of averaged polarizations. The coupled equations are discretized in space and time using a finite-difference time-domain method which provides a versatile platform to design complex structures and integrate diverse material models. We believe that our models are crucial tools enabling for the realization and optimization of optical limiting and all-optical switching systems.
\end{abstract}

\maketitle

\section{Introduction}
There has been a renewed interest in the optical community to protect optical sensors and more importantly human eyes from accidental or intentional harm caused by high intensity light sources. This is accomplished by optical limiters which have a low absorption at low light intensities and high absorption at higher intensities. There are multiple nonlinear processes that could be utilized for optical limiting including nonlinear scattering, absorption, refraction, and multi-photon absorption~\cite{ol_materials1999}. 
Materials exhibiting nonlinear absorption, specifically reverse saturable absorption (RSA), are amongst the most popular for optical limiters~\cite{ol_materials1999}. RSA materials provide high transmission (low absorption) at low intensity of incident light, and low transmission when the intensity of incident light becomes high.
Defined more rigorously, RSA occurs when the absorption cross-section of the excited states is larger than that of the ground states. RSA also requires the lifetime of the triplet state  to be relatively long, which is reflected directly in the efficiency and the saturation intensity of the optical limiter (OL). 
A wide range of materials that exhibit RSA includes metals \cite{ag_cu2004}, organic dyes \cite{rohdamineB2003, c60_1994}, clustered metal particles \cite{metal_cluster1990, silver2008}, synthetic compounds \cite{tetraphenylporphyrins1985}, carbon nanotubes \cite{carbon_nano2000}, and two-dimensional materials \cite{graphene2015,graphene_mos2_2013}.

Unlike other nonlinear phenomena, there are surprisingly few efforts to develop complete multiphysics models of optical limiting devices based on RSA with resonant nanostructured optical elements. Such models would allow for the significant improvement of the size, weight and power metrics of this class of OLs.
One of the most sought after requirements is to obtain RSA at normal radiation levels, which would greatly expand their area of applicability \cite{natmat2014, large2017} and open up new opportunities for obtaining high RSA materials at weak incoherent light.
Most RSA materials have extremely high saturation energies, defined as the energy required to reduce the linear transmittance to $1/e$ of its linear value. This limits the use of OLs based on those materials to very intense and/or very short laser pulses.  
Current efforts to optimize optical limiting properties are materials focused. Using material synthesis, chemical engineering and functionalization of materials to increase their triplet state life time have been widely explored to get enhanced RSA characteristics at sunlight power level \cite{natmat2014,large2017}.
We believe that a critical limiting factor hindering the optimization and enhancement of OLs on the device level is the lack of material models that accounting for the quantum mechanical origin of RSA and its relevant representation in the time domain. 
Once an accurate model of RSA is available, boosting the performance of OLs using the current advances in nanophotonics and micro- and nano-fabrication should be made substantially more achievable.
Thus far, there have been efforts to model saturable nonlinearities in the time domain using various methods such as a saturable harmonic oscillator \cite{varin2015saturable} as well as two-level systems for saturable absorbers \cite{scattering2001,mock2017}. For RSA, the physical mechanism causing the absorption is somewhat more involved and requires taking into account different time-scales as well as the absorption contributions of both the singlet and triplet states.

In this work, we propose an advanced time domain numerical simulation approach that models the RSA process as an atomic multi-level system in a 3D full-wave multiphysics framework. With this approach we can simulate the different relaxations taking place in the material and get an idea of how they affect the nonlinear optical material response. Additionally, we can also develop a deeper understanding of the excitation mechanisms, and then employ this comprehensive knowledge to optimize optical limiting devices and systems. Similar to the experiment-fitted time-domain kinetic models of gain media \cite{Trieschmann2011}, feedback from optical experiments should always be recommended for accurate material modeling in design and optimization of real-life RSA devices.
Advantages of our scheme include: (i) self-consistent description of the excitation and relaxations dynamics, coupled to (ii) 3D full-wave multiphysics environment with the geometry discretization that can handle an arbitrary structural complexity, and accurately model focused/structured beams that match experimentally-used sources, without resorting to paraxial or non-paraxial approximations, and finally, (iii) in-depth post-processing of the simulated data -- time resolved tracking of $E$-fields, polarizations and population kinetics --  getting physical insights into the system dynamics not otherwise available through experiment.

	Following this introduction, a particular topology of a multi-level atomic system for modeling RSA is introduced and explained with more details. Then, the numerical framework and results from the discretized models are discussed. Time dynamics of the system along with the effects of controlling parameters on the system optical response are studied. 
	Finally, the results from our model are compared to the classical Beer-Lambert law by fitting the corresponding absorption cross-sections  verifying that our model can reproduce experimentally measured RSA data.
	
\begin{figure}[!htb]\centering
	\includegraphics[scale=1]{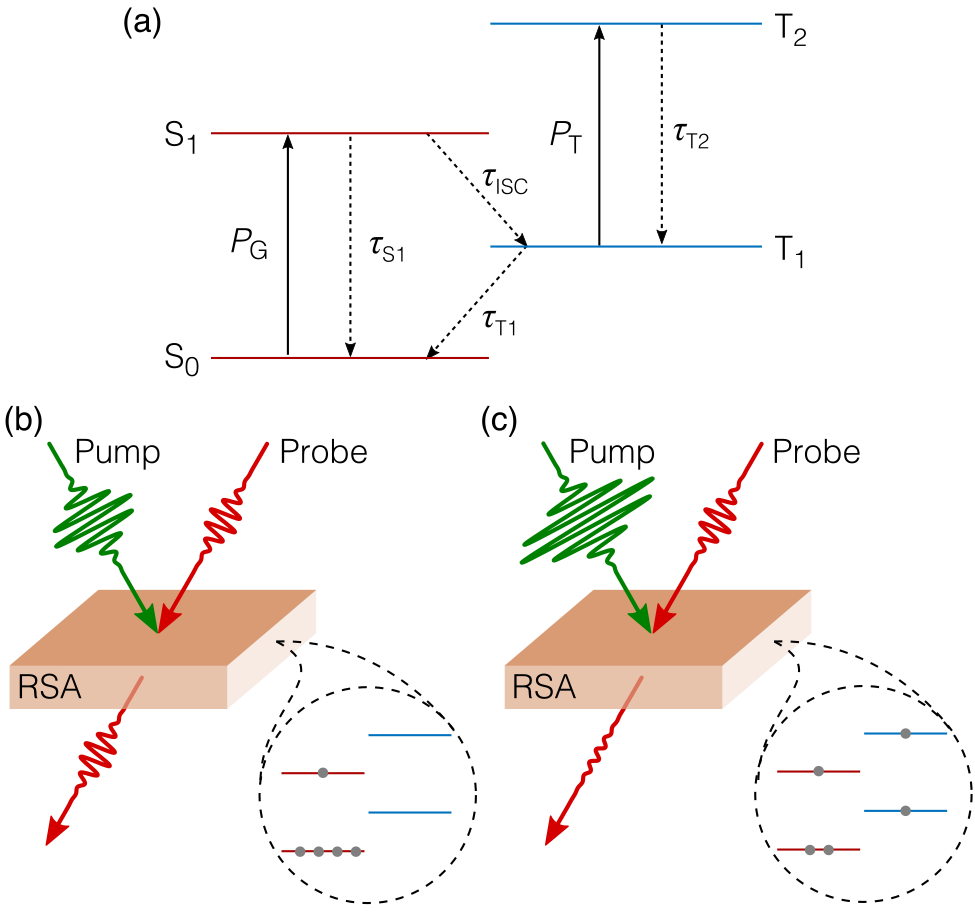}
	\caption{RSA mechanism. (a) Jablonski diagram of absorbing medium represented by a four-level atomic system. $S_{0,1}$ and $T_{1,2}$ represent the singlet and triplet states respectively. (b)~Linear regime: low-intensity pump with no sufficient population in the lower triplet state. (c)~Nonlinear regime: high intensity pump with absorption taking place in the triplet states leading to RSA behavior.
}\label{jablonski}
\end{figure}
     
\section*{RSA models} \label{models}
The band diagram of a material with RSA can be represented by an atomic four-level system with the Jablonski diagram shown in Fig. \ref{jablonski}(a). 
When light illuminates an RSA medium, the atom is excited from the  ground singlet state $\mathrm{S}_0$ to the first singlet state $\mathrm{S}_1$ by absorbing a photon of the incident light, Fig. \ref{jablonski}(b). The excited atom at $\mathrm{S}_1$ can return back to $\mathrm{S}_0$ by radiating the same photon or can exhibit an intersystem crossing (ISC) relaxation to the first triplet state $\mathrm{T}_1$. The atom at $\mathrm{T}_1$ can either relax to $\mathrm{S}_0$ or can be promoted to the second triplet state $\mathrm{T}_2$ by absorbing a second photon if the incident light intensity is sufficient, Fig. \ref{jablonski}(c). The additional absorption associated with the elevations of atoms from $\mathrm{T}_1$ to $\mathrm{T}_2$ is the origin of the RSA. It can be thought of as an additional channel for absorption in addition to the ground state absorption channel (singlet pair).
For the RSA to be efficient, a sufficient population of carriers should be transferred to $\mathrm{T}_1$, which requires efficient ISC and a long triplet state lifetime. Also, the absorption cross section of the excited triplet states must be larger than that of the ground singlet state.

\subsection{Beer-Lambert's law} \label{beer}
The Beer-Lambert law (BLL) is commonly used in spectroscopy and chemical analysis to relate the absorbance of a given absorbing material to its concentration in a host medium and the path length.  
The rate equations (RE) that model carrier kinetics at the different states are commonly employed to approximate the absorption coefficient of the materials \cite{c60_1994, eric1999, eric2002}. Using the absorption coefficient with the BLL, the intensity decay in uniformly absorbing samples can be found. 
Assuming three allowed relaxations: intersystem crossing from $\mathrm{S}_1$ to $\mathrm{T}_1$, relaxation form $\mathrm{S}_1$ to $\mathrm{S}_0$ and relaxation from $\mathrm{T}_1$ to $\mathrm{S}_0$. The RE governing the transitions inside the material are given by: 
\begin{subequations} \label{eqn:beer}
\begin{eqnarray}
\frac{\mathrm{d}N_\text{S0}}{\mathrm{d}t} &=&  \frac{N_\text{S1}}{\tau_\text{S1}} +\frac{N_\text{T1}}{\tau_\text{T1}} -N_\text{S0}\frac{\sigma_\text{G} I}{h \nu} \\
\frac{\mathrm{d}N_\text{S1}}{\mathrm{d}t} &=&  -\frac{N_\text{S1}}{\tau_\text{S1}} -\frac{N_\text{S1}}{\tau_\text{ISC}}  +  N_\text{S0}\frac{\sigma_\text{G} I}{h \nu} 
\\
\frac{\mathrm{d}N_\text{T1}}{\mathrm{d}t} &=&  \frac{N_\text{S1}}{\tau_\text{ISC}}  - \frac{N_\text{T1}}{\tau_\text{T1}} 
\end{eqnarray}
\end{subequations}
\noindent
where $N_\text{S0}$, $N_\text{S1}$, and $N_\text{T1}$ are the population densities in $\mathrm{S}_0$, $\mathrm{S}_1$, and $\mathrm{T}_1$, where for simplicity the time dependency of intensity in Eq. (\ref{eqn:beer2}) is omitted.
The intensity of the light inside the sample can then be calculated using:
\begin{equation}\label{eqn:beer2}
\frac{\partial I}{\partial z} = -(N_\text{S0}\sigma_\text{G} + N_\text{T1}\sigma_\text{T})I
\end{equation}
Using this model, the time dependent absorbance can be calculated and the propagation of light intensity inside the sample can be straightforwardly obtained with a controlled accuracy and high efficiency even for tightly focused beams \cite{eric1999}. This approach, however, doesn't provide a material model that can be used to simulate more involved optical devices with dispersive or nonlinear materials and other multiphysics complications. Also, this simple, yet efficient, methodology doesn't allow access to the electric \textbf{E} or magnetic \textbf{H} vector fields, local population kinetics close to the device nanostructured boundaries, and may not account for pumping dynamics following structured illumination. All these limitations are addressed with our full-wave multiphysics numerical framework that includes a multi-level atomic model.

\begin{figure}[!htb]\centering
	\includegraphics[scale=1]{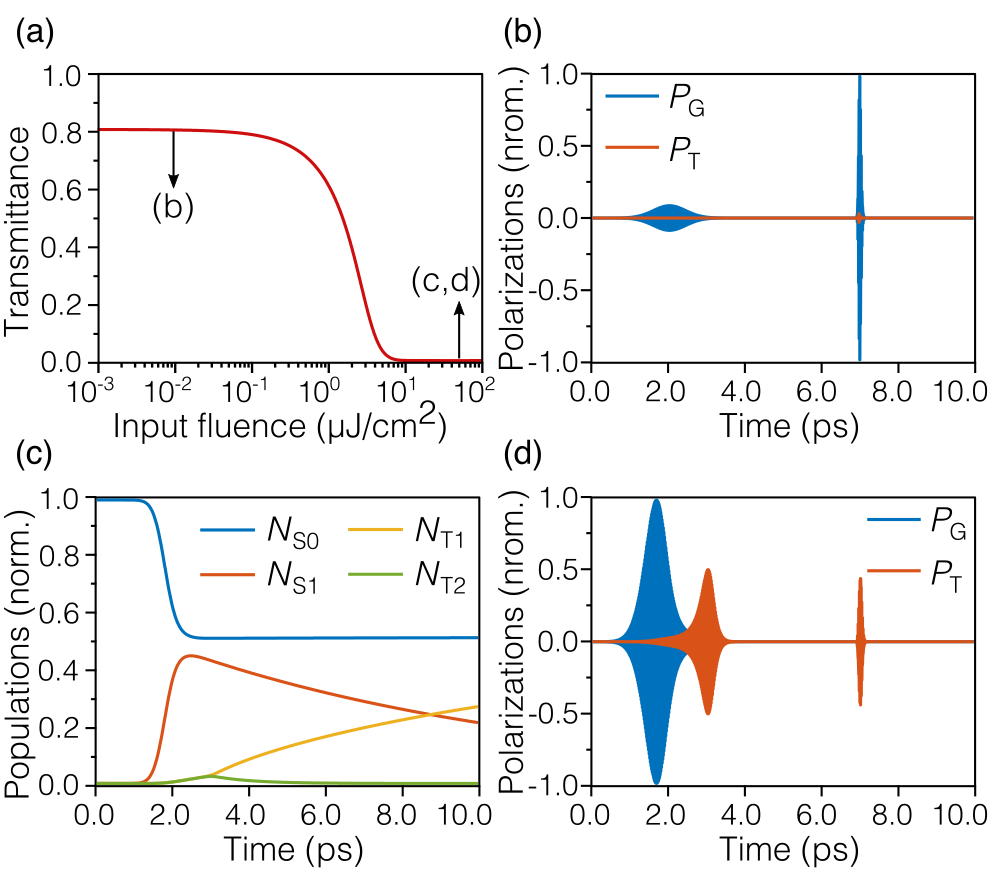}
	\caption{RSA behavior and time dynamics of four-level  system. (a) Transmittance versus input fluence of a 1-$\mu m$ film (b) Induced polarizations densities inside the RSA at 0.01 $\mu J/cm^2$ ,polarizations are governed by the ground state as the triplet state contribution in the linear regime is negligible (not shown). Nonlinear regime time dynamics for pump fluence of 50 $\mu J/cm^2$: (c) and (d) state densities and (d) polarizations.}
\label{fig:dynamics} 
\end{figure}

 \subsection{Four-level atomic system} \label{four}
As an example of a multi-level system, the atom or molecule of an RSA material is modeled using a four-level system with the Jablonski diagram shown in Fig. \ref{jablonski}(a).
The RE that govern all allowed transitions in a four-level system are given by: 
\begin{subequations} \label{4levels}
\begin{eqnarray}
\frac{\partial N_\text{S0}}{\partial t} &=&  \frac{N_\text{S1}}{\tau_\text{S1}} +\frac{N_\text{T1}}{\tau_\text{T1}} - \frac{1}{\hbar\omega_{0}}\textbf{E}\cdot  \frac{\partial \textbf{P}_\text{G}}{\partial t}\\
\frac{\partial N_\text{S1}}{\partial t} &=&  -\frac{N_\text{S1}}{\tau_\text{S1}} -\frac{N_\text{S1}}{\tau_\text{ISC}}  +  \frac{1}{\hbar\omega_{0}}\textbf{E}\cdot  \frac{\partial  \textbf{P}_\text{G}}{\partial t} 
\\
\frac{\partial N_\text{T1}}{\partial t} &=&  -\frac{N_\text{T1}}{\tau_\text{T1}} + \frac{N_\text{S1}}{\tau_\text{ISC}} + \frac{N_\text{T2}}{\tau_\text{T2}}- \frac{1}{\hbar\omega_{0}}\textbf{E}\cdot  \frac{\partial \textbf{P}_\text{T}}{\partial t}
\\
 \frac{\partial N_\text{T2}}{\partial t} &=&  -\frac{N_\text{T2}}{\tau_\text{T2}} +  
 \frac{1}{\hbar\omega_{0}}\textbf{E}\cdot  \frac{\partial \textbf{P}_\text{T}}{\partial t}.
\end{eqnarray}
\end{subequations}
The population densities in the lower and upper singlet and triplet states, respectively, are $N_\text{S0}$, $N_\text{S1}$, $N_\text{T1}$, and $N_\text{T2}$. The total population,  $N_\text{d} = N_\text{S0} + N_\text{S1}+ N_\text{T1} + N_\text{T2}$, is conserved at all times, where $N_\text{d}$ is the density of absorbing atoms. The lifetime of state $\mathrm{S_1}$ is $1/(1/\tau_\text{S1} + 1/\tau_\text{ISC})$ where $\tau_\text{ISC}$ is the intersystem crossing lifetime.  Lifetimes of the upper and lower triplet states are  $\tau_\text{T2}$ and $\tau_\text{T1}$. The term $\frac{1}{\hbar\omega_{0}}\textbf{E}\cdot  \frac{\partial \textbf{P}_{ij}}{\partial t}$ represents the stimulated emission and absorption between levels $|i>$ and $|j>$.

The induced macroscopic polarizations due to the transitions satisfy the Lorentz ordinary differential equation given by:
\begin{equation}\label{pol}
\frac{\partial^2 \textbf{P}_{ij}}{\partial t^2} + \gamma_{ij} \frac{\partial\textbf{P}_{ij}}{\partial t} + \omega_{0}^2\textbf{P}_{ij} = \kappa_{ij} (N_j - N_i)\textbf{E} 
\end{equation}

\noindent
where $\gamma_{ij}$ is the dephasing constant between levels $|i>$ and $|j>$. The energy spacing between the levels is $\Delta E_\text{S10}  \approx \Delta E_\text{T21} = \hbar \omega_0$ where $\omega_0$ is the frequency of the absorbed light. 
The excitation term in Eq. (\ref{pol}) is proportional to product of the local electric field and the difference in populations between the upper and the lower levels of this transition with a proportionality constant $\kappa_{ij}$ where $ij \in \{\text{G}, \text{T}\}$. The coupling coefficient $\kappa_{ij}$ represents the dipole matrix element between levels $|i>$ and $|j>$ and is given by 
$\kappa_{ij} = 2\omega_{ij}|\mu_{ij}|^2/\hbar = 6\pi\epsilon_0 c^3/\omega_{ij}^2 \tau_{ij}$,
where $\epsilon_0$ and $c$ are the free space permittivity and speed of light, respectively, and $\tau_{ij}$ is the decay time constant between levels $|i>$ and $|j>$ \cite{taflove2004}. 
The polarization densities are then coupled to the Maxwell equations through the electric flux,  
\begin{equation}\label{hfield}
\nabla \times \textbf{H}(t, \textbf{r}) = \epsilon_0 \epsilon_h \frac {\partial{\textbf E}(t, \textbf{r})} {\partial t} + \frac{\partial{\textbf P}(t, \textbf{r})}{\partial t}
\end{equation}
\begin{equation}\label{efield}
\nabla \times \textbf{E}(t, \textbf{r}) = -\mu_0 \frac {\partial{\textbf H}(t, \textbf{r})} {\partial t},
\end{equation}
\noindent
and Eqs. (\ref{4levels})-(\ref{efield}) are solved simultaneously using a full-wave solver on a staggered Yee grid by applying the classical central finite-difference approximation to the spatial and time derivatives.\cite{taflove2004} The time dependent electromagnetic fields are recorded and converted to frequency domain using fast Fourier transform to obtain the transmission and reflection, and finally the absorption.    

\section{Numerical Details and Results}\label{results}
We model a 1-$\mu m$ film using our proposed four-level atomic model. 
The life-time and scattering rate parameters of the model are: $\tau_\text{T2}=1~ps$, $\tau_\text{T1}=300~ns$, $\tau_\text{S1}=1~ns$, $\tau_\text{ISC}=10~ps$, and $\gamma_\text{G} = \gamma_\text{T} = 10^{14}~Hz$.
The density of the absorptive molecules is 0.3 $mM$ (1.806$\times 10^{17} cm^{-3})$. The sample is illuminated with two pulses separated by a short delay of 5 $ps$: a strong 1-$ps$ pump pulse followed by a weak 50-$fs$ probe. The probe fluence is fixed at $100 nJ/cm^2$ while the fluence of the pump is varied from $10^{-3}~\mu J/cm^2$ to $10^2~\mu J/cm^2$ to study the nonlinear transmission and absorption of the thin RSA film. 
The wavelengths of both the pump and the probe are 532~$nm$.
Figure \ref{fig:dynamics} shows the RSA behavior and time dynamics of the thin film. In Fig. \ref{fig:dynamics}(a), the transmission is linear (constant transmittance) at low input (pump) fluence and nonlinear at higher values, the saturation fluence for the RSA film is around $2.5~\mu J/cm^2$ in this example. The time dynamics of the system at two pump fluences of $10^{-2}~\mu J/cm^2$ and $50~\mu J/cm^2$ are depicted in Fig \ref{fig:dynamics}(b)-(d). 
At $10^{-2}~\mu J/cm^2$, well into the linear regime, the normalized ground state population density is almost unity during all times and consequently the carrier build-up in the upper states is negligible (not shown). The time evolution of the macroscopic polarization densities of the ground ($P_\text{G}$), and excited triplet ($P_\text{T}$) states are shown in  Fig. \ref{fig:dynamics}(b) and, as expected, $P_T$ is almost negligible in this case. All polarizations are normalized to $P_\text{G}$ for clarity. Further, in the nonlinear regime with pump fluence of $50~\mu J/cm^2$, the carrier dynamics, Fig. \ref{fig:dynamics}(c), show depopulation of the ground state and increased build-up of triplet state density, hence, $P_\text{T}$ starts to contribute to the material response, Fig. \ref{fig:dynamics}(d). We can clearly notice that the response of the system to the probe (around 7~$ps$) is completely dominated by the $P_\text{T}$  which explains the enhanced absorption in this case. 

\begin{figure}[!htb]
	\includegraphics[scale=01]{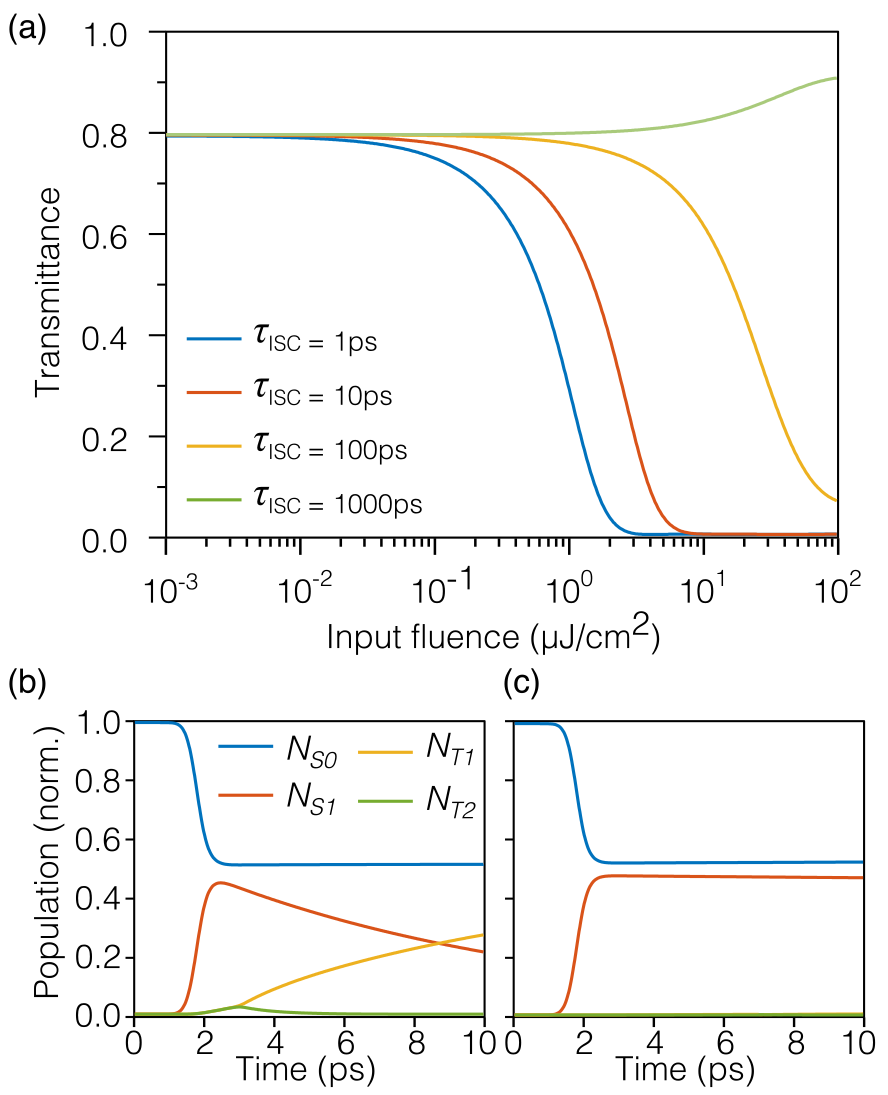}\centering
	\caption{Effect of triplet quantum yield  on the RSA efficiency. (a) Transmittance versus pump fluence at four different values of $\tau_{ISC}$: 1, 10, 100 and 1000 $ps$ corresponding to  $\Phi_\text{T} $ of 0.999, 0.99, 0.909, and 0.5, respectively. (b), and (c) the population time dynamics at 10, and 1000 $ps$ at pump energy of 50 $\mu J/cm^2$. }

\label{fig:yield} 
\end{figure}

Crossings of molecules from $\mathrm{S}_1$ to $\mathrm{T}_1$ requires undergoing a spin conversion in a process called intersystem crossing \cite{principles_of_spectroscopy2008}. The efficiency of the intersystem crossing is determined by the triplet quantum yield $\Phi_\text{T} = (1/\tau_\text{ISC}+1/\tau_\text{S1})/\tau_\text{ISC} $. The faster the intersystem rate ($1/\tau_\text{ISC}$), the higher the triplet quantum yield. 
Here, we study the effect of $\Phi_\text{T}$ on the system response. We study the same sample as before while varying the intersystem crossing lifetimes to 1, 10, 100 and 1000 $ps$ corresponding to $\Phi_\text{T}$ of 0.999, 0.99, 0.909, and 0.5, respectively. Transmittance versus input fluence at the four different cases are depicted in Fig. \ref{fig:yield}(a). We notice that for $\tau_\text{ISC}$ of 1, 10, and 100  $ps$, the material shows RSA behavior with increasing saturation fluence. However at $\tau_\text{ISC}$ of 1 $ns$, the material acts as a saturable absorber. This matches the system dynamics very well. As seen in Fig. \ref{fig:yield}(b), at $\tau_\text{ISC}$ of 10~$ps$, the efficiency of the triplet quantum yield is high enough to allow for a sufficient population transfer to $T_1$ unlike at $\tau_\text{ISC}$ of 1~$ns$, Fig. \ref{fig:yield}(c), where the population density at $T_1$ is almost negligible and as a result, the carrier density at $S_1$ starts to saturate leading to saturation of system absorption and hence to increased transmission (see the green line in Fig. \ref{fig:yield}(a)). 

\begin{figure}[!htb]
	\includegraphics[scale=1]{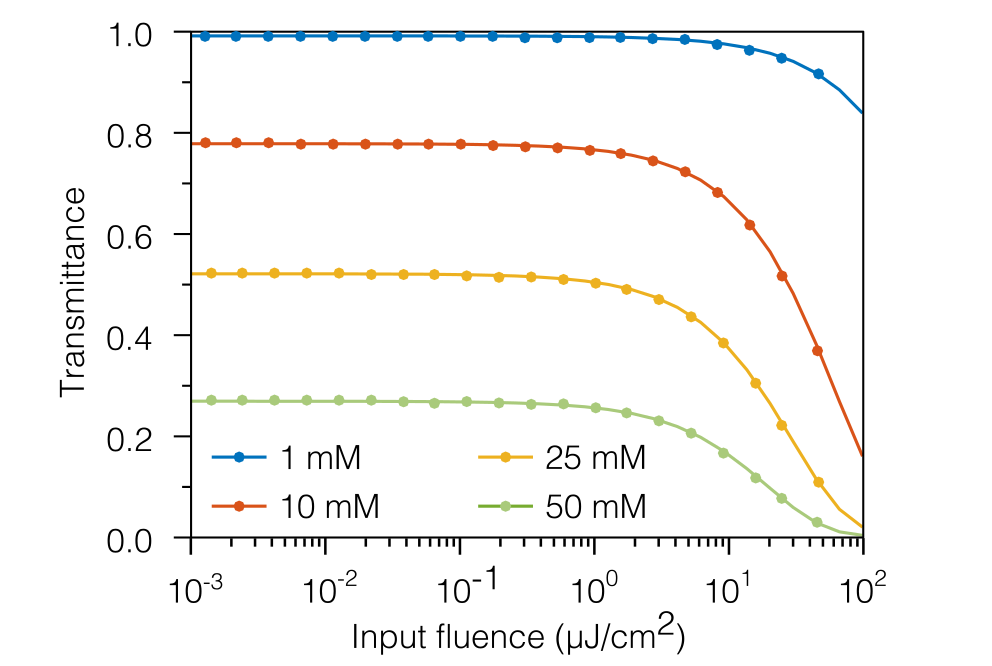}\centering
	\caption{Transmittance versus pump fluence using full-wave atomic system simulations (solid lines) vs BLL (dotted lines) at different molecule densities.}
    
\label{fig:beer} 
\end{figure}

Next, we compare the results of a 1-$\mu m$-thin RSA film with varying $N_\text{d}$ using a four-level atomic model and the classical BLL. The lifetimes of the films are $\tau_\text{T2}=1~ps$, $\tau_\text{T1}=1~\mu s$, $\tau_{S1}$=30 $n s$, and $\tau_\text{ISC}=0.1~ns$. The same lifetimes are used in both simulations while changing the absorption cross-sections to fit the two models. Figure \ref{fig:beer} shows very good agreement between the full-wave simulation (solid lines) and the BLL equations (dotted lines) calculations. Generally the BLL has been successfully used to fit experimental results \cite{eric1999,eric2002} of uniform samples. Having an excellent match with the BLL implies the ability of our model to fit and further utilize experimentally measured RSA transmission data. From Fig. \ref{fig:beer}, the saturation intensity as well as the linear transmittance increases with decreasing absorption molecular densities which suggests the use of thinner films with larger $N_\text{d}$ in optical limiters device engineering.

\section{Summary and Outlook} 
In conclusion, a 3D full-wave analysis of the RSA devices with the ability to model light-matter interaction based on rate equations that could include non-paraxial beams, complex structures and diverse material composition has been presented and validated. The results of the proposed model agree well with the classical Beer-Lambert equation used to fit experimental RSA data while also providing a comprehensive physical understanding of the fields' and carriers' dynamics within the absorbing material.
Using our technique, the numerical modeling and optimization of  micro- and nano-structured devices are made possible which opens up new opportunities to explore high-sensitivity, low-threshold optical limiters beyond homogeneous RSA material limitations. 
 
\textbf{Funding.} Authors acknowledge the financial support by DARPA/DSO Extreme Optics and Imaging (EXTREME) Program, Award HR00111720032.

\bibliography{ref.bib}

\end{document}